# Heterotwin Zn$_3$P$_2$ superlattice nanowires: the role of indium insertion in the superlattice formation mechanism and their optical properties.

Simon Escobar Steinvall[1*], Lea Ghisalberti[1*], Reza R. Zamani[2*], Nicolas Tappy[1], Fredrik S. Hage[3,4], Elias Stutz[1], Mahdi Zamani[1], Rajrupa Paul[1], Jean-Baptiste Leran[1], Quentin M. Ramasse[3,5], W. Craig Carter[1,6], Anna Fontcuberta i Morral[1,7]

* Equal contribution

1 Laboratory of Semiconductor Materials, Institute of Materials, École Polytechnique Fédérale de Lausanne, 1015 Lausanne, Switzerland

2 Centre Inderdiscplinaire de Microscopie Électronique, École Polytechnique Fédérale de Lausanne, 1015 Lausanne, Switzerland

3 SuperSTEM Laboratory, SciTech Daresbury Campus, Keckwick Lane, Warrington WA4 4AD, United Kingdom

4 Department of Materials, University of Oxford, Oxford OX1 3PH, United Kingdom

5 School of Chemical and Process Engineering and School of Physics and Astronomy, University of Leeds, Leeds LS2 9JT, United Kingdom

6 Department of Materials Science and Engineering, Massachusetts Institute of Technology, Cambridge, Massachusetts 02139, USA

7 Institute of Physics, École Polytechnique Fédérale de Lausanne, 1015 Lausanne, Switzerland

## Abstract

Zinc phosphide (Zn3P2) nanowires constitute prospective building blocks for next generation solar cells due to the combination of suitable optoelectronic properties and an abundance of the constituting elements in the Earth's crust. The generation of periodic superstructures along the nanowire axis could provide an additional mechanism to tune their functional properties. Here we present the vapour-liquid-solid growth of zinc phosphide superlattices driven by periodic heterotwins. This uncommon planar defect involves the exchange of Zn by In at the twinning boundary. We find that the zigzag superlattice formation is driven by reduction of the total surface energy of the liquid droplet. The chemical variation across the heterotwin does not affect the homogeneity of the optical proerties, as measured by cathodoluminescence. The basic understanding provided here brings new perspectives on the use of II-V semiconductors in nanowire technology.

## Introduction

Filamentary crystals, also known as nanowires, have provided additional design freedom in the elaboration of materials with desirable properties.[1–4] This arises from the possibility of engineering the crystal phase, material composition, and for the possibility of expanding the structure in three dimensions.[3,5,6] Among the design opportunities, the composition or structure of nanowires can be arranged periodically in the form of superlattices.[6–9] The periodicity of the superstructure modulates both the electronic and phonon (vibrational) states, depending on the magnitude of the period.[10–13] Semiconductor superlattices find applications in the optoelectronic and thermoelectric arena.[14–17] In thin films, the materials combinations are mostly restrained due to lattice-mismatch and thermal expansion conditions. Superlattice nanowire structures circumvent these limitations, and have been achieved by modulating the composition, crystal phase, and crystal orientation through rotational twins.[5–8]

     Twin superlattices (TSLs) in semiconductors were predicted by Ikonic *et al.* in 1993.[18] More recently, they were implemented in nanowire form, first in Al$_2$O$_3$ and ZnSe, and subsequently in InP.[19–21] These TSLs have been obtained mainly by the vapour-liquid-solid (VLS) method in which a nanoscale liquid droplet preferentially collects the growth precursors.[22] In addition to a periodic arrangement of twins, these nanowire superlattices adopt a characteristic zigzag morphology with alternating (111)A/B facets in the case of zincblende nanowires.[7,8] According to Algra *et al.*, twin formation is determined by energy minimisation involving the stability of the droplet and the surface energy of the facets as a function of their polarity –(111) a or B-.[8]

     Zinc phosphide, Zn$_3$P$_2$, has recently attracted attention as a compound semiconductor made of elements that are abundant in the Earth's crust with optoelectronic properties suitable for photovoltaic applications.[23–28] Zn$_3$P$_2$ has been obtained both in the form of bulk crystals[29,30], thin films[25,26,31], and nanostructures[9,24,32–34]. Zinc phosphide based solar cells with an efficiency of up to 6% have been reported.[30]

This value is still well below the theoretical limit (>30%), illustrating the improvement potential of this material.[35,36]

The synthesis of $Zn_3P_2$ nanowires can follow the VLS and the vapour-solid mechanisms, with In, Sn, and Au as catalysts.[9,24,32–34,37,38] $Zn_3P_2$ nanowires adopt various morphologies depending on the fabrication method and/or growth conditions, including a zigzag superlattice.[9,24,32–34] In contrast with III-V compound semiconductors, $Zn_3P_2$ exhibits a centrosymmetric tetragonal structure, and thus also non-polar facets and main crystal symmetry directions.[29] Consequently, all side facets of $Zn_3P_2$ zigzag nanowires are always Zn-terminated.[27,33] This means that the mechanism through which $Zn_3P_2$ obtains a zigzag morphology is inconsistent with the model proposed based on III-Vs.[8]

In this paper we reveal the nature of the defects leading to the zigzag structure using aberration-corrected and analytical scanning transmission electron microscopy (STEM). In addition, we explain the formation mechanisms based on simulations of the surface energetics of the droplet as a function of the nanowire cross-section. Finally, we outline the consequences of this periodic structure for the optical functionality through cathodoluminescence spectroscopy (CL).

## Experimental

The $Zn_3P_2$ nanowires were epitaxially grown in a Veeco GENxplor molecular beam epitaxy (MBE) system on InP (100) substrates. They were grown through In catalysed VLS, with the In originating from the substrate.[24] The analysed samples were grown at a manipulator temperature of 250°C and a V/II ratio of 1.15 or 1.45 for four hours, with additional details on the growth in [24]. The nanowires were transferred to copper TEM grids with holey carbon by scraping the grid on the growth substrate for TEM studies, and were used as grown for CL studies.

Scanning electron microscopy (SEM) images were acquired using a Zeiss Merlin FE-SEM equipped with a Gemini column. The operating conditions were an acceleration voltage of 3 kV and a beam current of 100 pA. An in-lens secondary electron detector was used for the imaging.

Droplet simulations were performed using the Surface Evolver software[39], which computes minimised surface energy by optimising shapes given constraints and wetting angles. We implement the interfacial energy of the liquid-solid interface by means of the Young's equation with a contact angle of 43° when the triple line is unconstrained. The certical axis in Fig 2c is the total energy divided by $L^2$, with L being the average length of the sides, and the difference between the solid-liquid and the solid surface tensions. The average length of the side remains constant with a value of 1.57. Regarding the geometrical constraint, the triple line is not pinned to the edge, but is left free to move inside the polygon. To build the polygon, centred at the origin, we define an equation for each side of the hexagon through the lat and shrink parameters, illustrated in the SI. While the lat parameter is fixed at 0.55 and defines the apothem of the reference hexagon, the shrink parameter is variable controlling the shape of the constraint since it represents the normal between the facet centroid and the selected facet. By varying the shrink parameter from -0.25 to 0.25, we can reproduce the evolution of the nanowire cross-section from left oriented triangle to right oriented triangle (HT1 & HT2), passing through the hexagonal geometry at shrink equal to 0. To compare the effect of the volume we performed the simulations for three different droplet volumes: 0.125, 0.225, and 0.325 with dimensions of $L^3$.

Aberration-corrected bright-field/high-angle annular dark-field (BF/HAADF) STEM images and electron energy loss spectroscopy (EELS) maps were collected on a STEM-dedicated Nion microscope (US100MC) operating at 60 kV. The Nion UltraSTEM 100MC HERMES is equipped with a C5 Nion probe corrector (full correction up to 6-fold astigmatism C5, 6) and a UHV Gatan Enfinium ERS spectrometer optimised for high energy resolution with high-stability electronics. The microscope is equipped with a cold-field emission gun (C-FEG), having an energy spread of 0.35 eV. The beam convergence semi-angle was 31.5 mrad and the EEL spectrometer entrance aperture semi-angle was 44 mrad. Image detector angles were 0-14 mrad (BF) and 100-230 mrad (HAADF). To minimise contamination, the specimens were baked prior to insertion at 130 °C in vacuum (~$10^{-6}$ Torr), and the microscope column is maintained at ultrahigh vacuum (UHV). The denoising of STEM-EELS datasets was done using the MSA plugin for Gatan's Digital Micrograph suite, commercially available from HREM research.[40] Example spectra are shown in the SI, and the 443 eV and 1020 eV peaks were used for EEL mapping of In and Zn, respectively. Further imaging was also performed in a FEI Titan Themis 60-300 kV TEM operating at 200 or 300 kV. The machine is equipped with a field emission gun (X-FEG), a monochromators, two aberration correctors (one pre-specimen probe-corrector, and one post-specimen image corrector), and a Fischione HAADF detector. The collection angles are typically 85-200 mrad for HAADF-STEM images. The BF and HAADF images were denoised using radial a Weiner filter. Viewing direction illustrations were created in Mathematica.

An Attolight Rosa setup equipped with an Andor Newton 920 Si-CCD was used for CL measurements. It was operated at room temperature with an acceleration voltage of 3 kV, a beam current of

<1 nA, and an exposure time of 50 ms per pixel. The nanowires were mounted on a stage with 20° tilt. Denoising of the hyperspectral maps was done through principal-component analysis (PCA) using the Hyperspy Software.[41] Peak fitting was done after data treatment based on the approach described in ref. [42].

## Results and Discussion

**Electron Microscopy**

Figure 1a shows a representative secondary electron SEM images of a typical zigzag $Zn_3P_2$ nanowire. These nanowires grow perpendicular to (101) with side facets belonging to {101}.[24,33] While the lateral facets in III-V superlattices, (111)A and (111)B, can exhibit different polarities, this is not the case for $Zn_3P_2$ (101) facets as they are all Zn-terminated.[7,8,27] The supposed difference between the lateral facets has been cited as one of the driving forces for the formation of TSL nanowires. This argument cannot be applied to this case as $Zn_3P_2$ does not display polar facets. Before discussing the mechanism, we disclose the nature of the interface dividing the zigzag regions in the nanowire.

Figure 1b shows an aberration-corrected BF-STEM image of a zigzag nanowire ([111] zone axis), revealing a "twin-like" planar defect separating different segments that are mirrored. This structural defect is akin to the twin planes in the III-V TSL nanowires: it interfaces two segments of the nanowire that appear to be rotated 180° around the nanowire growth axis –the (101) plane-, which is the most common twin plane in tetragonal systems.[43]

Figure 1c shows the HAADF-STEM image of a region equivalent to that shown in Figure 1b, viewed along the other major zone axis, [-101]. The bottom insets corresponds to "close ups" of the interfaces, identifying sets of trimers (Zn – blue, P – red) at each side of the defect. The topmost insets show a three-dimensional model of the structure, highlighting the viewing direction of the respective zone axes. While the defect is easily discerned by the increase in intensity and break in periodicity in Figure 1c, the structure is not perfectly mirrored along this zone axis. Instead, we observe a translation of the top crystal by a (400) plane along the <100> direction perpendicular to the zone axis. Furthermore, as corroborated below, the interfacial defect extends to more than one monolayer, and is of a different composition than the neighbouring segments. Thus, the two rotated crystals are separated, not sharing any crystal lattice points, and consequently this defect cannot be considered a standard twin.

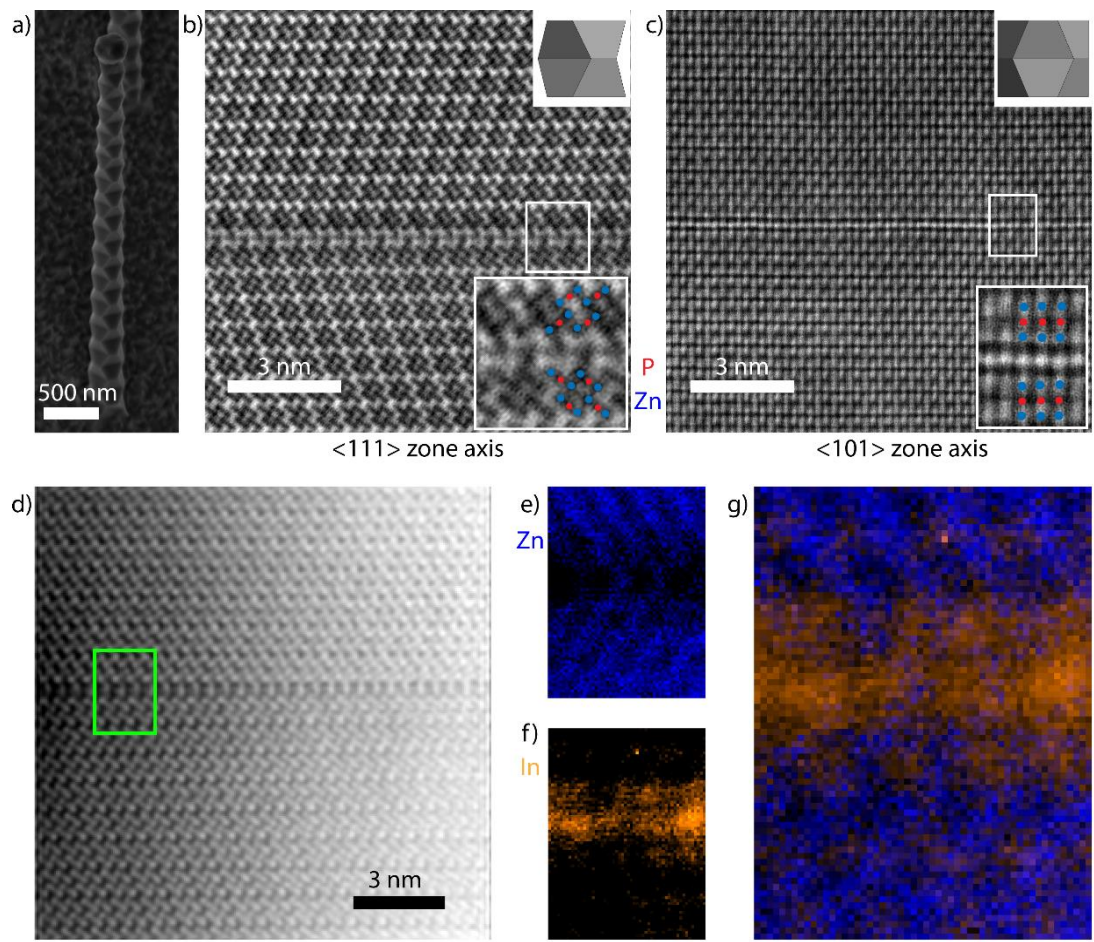

**Figure 1.** a) Secondary electron SEM image of a superlattice nanowire. b) Aberration corrected BF-STEM image taken along a [111] zone axis in the region around the zigzag interface with an inset illustrating the viewing direction. c) Aberration corrected HAADF-STEM image taken along a [101] zone axis in the region around the zigzag interface. The top inset illustrates the viewing direction and the bottom inset displays a "close up" on the interface. d) HAADF-STEM image ([111] zone axis) of the region where the EELS maps were acquired (green). e-g) Core-loss EELS maps of Zn (blue - e), In (orange - f), and the combination (g), showing the localised presence of In in the region around the stacking fault.

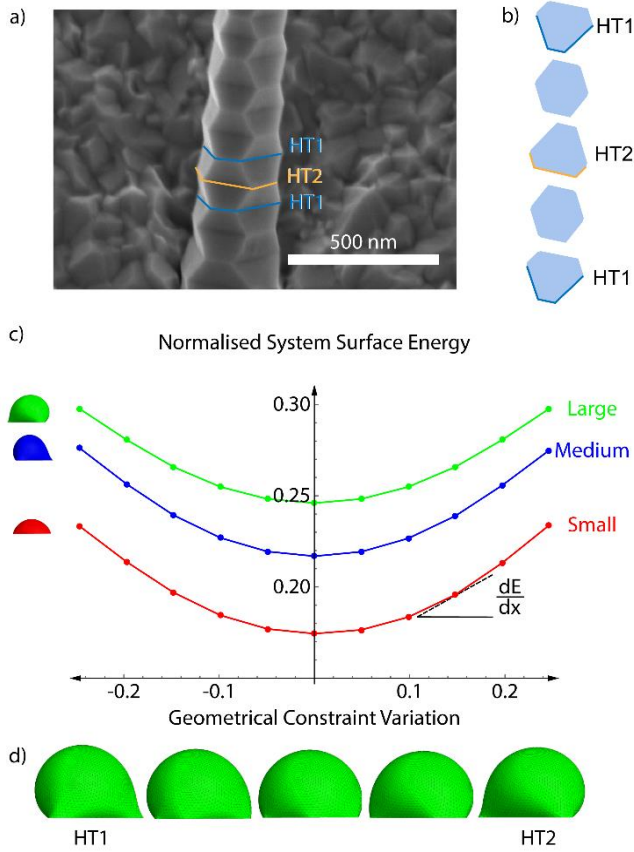

**Figure 2.** a) Secondary electron SEM image of a zigzag $Zn_3P_2$ nanowire b) schematic of the development of the cross-section of the nanowire as a function of the nanowire's growth axis, c) Results of the computation of the system (droplet and nanowires top facet) normalised surface energy (normalisation explained in the text) as a function of the variation of the geometry of the NW's top facet for three different referential droplet volume: small (red), medium (blue) and large (green); d) 3D sketch of the large droplet morphology upon the zigzag period.

We not turn to the study of the nature of the interface separating the twinned regions by providing data on the chemical composition. For this, we took the core-loss EELS of the "twin" and the adjacent regions –marked with a green square in Figure 1d, resulting in chemical maps with atomic-scale resolution.[44,45] Figure 1e-g display the resulting chemical mapping of Zn (blue – e), In (orange – f), and combined (g). P mapping did not provide the same resolution, and a constant signal was observed throughout. The maps reveal the presence of In at the interface and the neighbouring layers, accompanied by a drop in the Zn content. The presence of In is consistent with the intensity analysis in the HAADF-STEM images, especially from the [-101] direction (Figure 1c). The interface there appears slightly brighter than the rest, suggesting the presence of a heavier element, i.e. In. Given the chemical inhomogeneity across the boundary, the defect should rather be identified as a heterotwin.[46]

The utilisation of a chemical heterogeneity at the boundaries or planar defects have been reported in Al/TiN composites and in doped II-VI compounds such as ZnO. In the case of Al, N-terminated TiN lowers the formation energy of twins, and they provide significantly improved mechanical properties.[46,47] In ZnO, trivalent metals such as Al, Fe, Ga, or In have shown to precipitate at the interface of inversion domain boundaries.[48] The presence of trivalent metals in II-VI defects modifies the bonding coordination from four in a tetrahedral fashion to eight in an octahedral one, causing the polarity inversion.[49–52] However, in the case considered here the defect cannot be classified as an inversion domain boundary as there is no polarity inversion associated with it. In the following section we discuss the mechanism by which $Zn_3P_2$ forms a zigzag structure via a heterotwin.

**Zigzag Mechanism**

To investigate the driving forces prompting the regular insertion of heterotwins, we analysed the heterotwin periodicity along the nanowire length, $x\ (nm)$, and as a function of $W$, an approximation of the cross-sectional apothem. The trend was observed in multiple nanowires, while the equation is based on the high-resolution TEM image shown in Figure S3. Similar to III-V TSL nanowires, the heterotwin periodicity in $Zn_3P_2$ nanowires depends on their diameter.[7,53] The zigzag morphology results in the width being a periodic function with an amplitude with linear decay:

$$W(x) = \left(W_0 - \frac{W_0 - \frac{W_0}{2}}{2} - \frac{x}{\tan\frac{\pi}{2.02}}\right)\left(1 + \frac{1}{3}\cos\frac{2\pi x}{h_0 e^{-2\times 10^{-5}x}}\right) \quad (1)$$

Where $W_0$ is the initial width and $h_0$ is the distance between the initial segment separation. Equation 1 shows that the heterotwin interdistance reduces with the reduction of the nanowire diameter, in agreement with studies on non-tapered superlattice ZnSe and GaAs nanowires.[20,53] In particular, the term $\pi/2.02$ corresponds to 89°, i.e. the measured tapering angle characterising the reduction of the nanowire's diameter. The origin of the tapering is explored in detail in ref. [24].

Regarding possible explanations for the heterotwin formation mechanism, previous studies argued that twins in a zigzag structure form to minimise the nanowire surface energy.[8,54] The argument is reasonable for compound semiconductor nanowires exhibiting facets with different polarities and thus different surface energies. However, $Zn_3P_2$ is not polar. The basic structural unit consists of symmetric Zn-P-Zn trimers, instead of asymmetric cation-anion dumbbells such as In-P in InP, and its centrosymmetric crystal structure.[29,48,55] Thus, all facets in the zigzag structure are equivalent.[27] This means we cannot reasonably attribute the instigation of the $Zn_3P_2$ twinning process to nanowire surface energy minimisation alone. As discussed below, deformation of the liquid droplet during growth provides a more compelling argument.

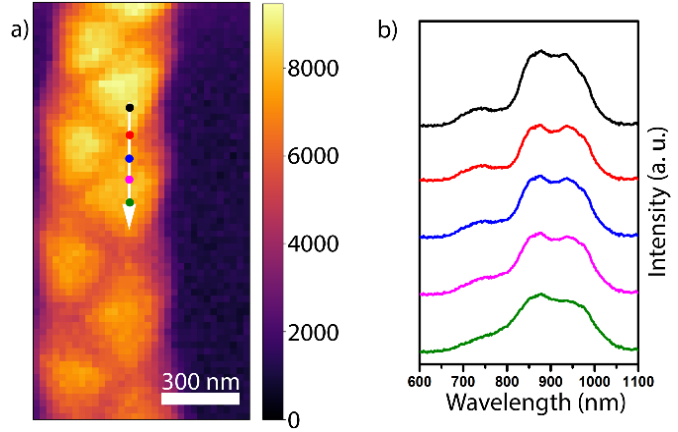

**Figure 3.** (a) High-resolution panchromatic CL intensity map of a zigzag nanowire with the white arrow indicating the linescan of the extracted spectra in (b), of which the second and fourth from the top were extracted from regions on top of the heterotwin.

We studied the droplet stability as a function of the volume and underlying cross-section of the nanowire, which varies during the zigzag formation as depicted in Figure 2a-b. To this end, we computed the surface energy of the liquid droplet and the interface with the nanowire using the finite-element based software Surface Evolver.[39] We used a Young angle of 43°, which is the experimental value found ex-situ.[56] Young's angles differing from 43° do not change the conclusion regarding alternating stability, the only modify the threshold for heterotwin formation.

Figure 2c illustrates the evolution of the surface energy of the droplet plus the liquid-solid interface during one zigzag cycle. We include the curves for three relative values of droplet volumes, which all follow a similar trend. The total surface energy has been normalised by the average length of the sides, constant throughout all simulations, by the surface energy of a floating sphere with an identical volume, and by the surface tension of the liquid-vapour interface. Due to the pinning of the liquid at the edge of the faceted nanowires, an increase in the liquid volume results in the increase of the apparent contact angle.[56] The normalised surface energy increases with the liquid volume due to the expansion and deformation of the liquid surface at the edges. The lowest normalised surface energy corresponds to the configuration with hexagonal nanowire cross-section, where the droplet is the least deformed. The normalised surface energy increases parabolically (to a second order approximation) with the modification of the liquid-solid interface area. The slope of the curve increases with the droplet deformation. A representative display of the predicted droplet shape during the zigzag process is shown in Figure 2d. It shows that the deformation is larger at the corners of the cross-section. In addition, the local and overall deformations are the largest when the cross-section is the closest to a triangular shape. This increasing deformation with the deviation from a hexagonal cross-section explains the increase in the normalised surface energy.

The introduction of a heterotwin constitutes a mechanism to stop the increase in surface energy during the nanowire elongation caused by the droplet's deformation.[53,57] As the energy required to form a heterotwin ($E_{HT}$) is fixed, there is a critical geometry of the nanowire's cross-section after which it is energetically favourable to insert a heterotwin rather than to continue increasing the total normalised surface energy. The probability of creating a heterotwin along the nanowire axis, $P(x)$, should thus increase with the normalised surface energy of the system, $\gamma(x)$, as:

$$P(x) \sim \frac{\gamma(x)}{E_{HT} A(x)} \qquad (2)$$

After formation of a heterotwin, the normalised surface energy decreases with nanowire elongation due to the change in the facet orientation and return towards a hexagonal cross section.

**Optical Properties**

Previous studies on $Zn_3P_2$ indicate that it exhibits a direct bandgap at ~1.5 eV, which is close to the ideal bandgap for the highest efficiency of single junction solar cells.[58] In our recent work, we demonstrated that zigzag $Zn_3P_2$ nanowires luminesce at 1.43 eV at cryogenic temperatures, which is relatively close to the expected value of the bandgap.[24] We have observed that in $Zn_3P_2$ nanowires with a square cross-section, luminescence can vary with the relative stoichiometry between the Zn and P.[24] Given the composition variation at the heterotwin in the zigzag nanowires, the question arises of whether the optical properties vary at these points. To ascertain their potential influence, we performed CL on zigzag nanowires at room temperature. The experimental conditions were chosen as to decrease the diffusion length and allow for higher spatial resolution measurements compared to previous studies.

Figure 3a shows a panchromatic map of the CL emission. We observe a ~10% variation in the emission intensity around edges and on the outward facet orientation compared to that of the inward orientation of the zigzag structure. We attribute this fluctuation to the variations in electron-beam excitation as a function of the morphology (see SI for more details). Detailed spectra along one oscillation of the zigzag morphology indicated by the arrow in Figure 3a are shown in Figure 3b. The spectra along and on each side of the heterotwin are qualitatively extremely similar. The similarity of the spectra along the zigzag structure could be due to the carrier diffusion length being larger than the excitation volume.[58] We observe to main peaks centred around 869 and 950 nm (1.43 eV and 1.30 eV, respectively), which correspond to sub-bandgap emission, potentially caused by the incorporation of indium in the bulk.[59] First principle simulations should be performed to confirm this hypothesis. We also distinguish a third weaker peak centred around 748 nm (1.66 eV). This peak could correspond to emission from the $\Gamma^2$ transition 260 meV above the bandgap, available due to the high energy excitation of CL.[58,60]

**Conclusions**

In conclusion, we have demonstrated that the $Zn_3P_2$ superlattice nanowires do not form through regular twinning, as observed in their III-V analogue. Instead, an In-rich heterotwin is formed, as shown through EELS mapping and aberration-corrected STEM imaging, which facilitates the rotation of the crystal structure between segments through the inset of a separate material. Furthermore, we developed a model to explain the onset of heterotwin formation. Based on the non-polar nature of $Zn_3P_2$ we could tie the model solely to the constraints posed by the droplet shape as a function of the nanowire cross-section. Characterisation of the emission through room temperature CL shows no effect of the heterotwins on the functional properties.

**Conflicts of interest**

There are no conflicts to declare.

**Acknowledgements**

S. E. S., E. S., M. Z., R. P., J. L., and A. F. i M. were supported by the SNSF Consolidator grant BSCGI0-157705. N. T. and A. F. i M. were supported by SNSF grant 20B2-1 17668. L.G., W. C. C., and A. F. i M. were supported by SNSF via project 200021_169908 and SCR0222237. R. R. Z. was supported by EPFL-CIME. F. S. H., Q. M. R., and the SuperSTEM Laboratory, the U. K. National Research Facility for Advanced Electron Microscopy, were supported by the Engineering and Physical Sciences Research Council (EPSRC).

**Bibliography**

1  J. Arbiol, M. de la Mata, M. Eickhoff and A. F. i Morral, *Mater. Today*, 2013, **16**, 213–219.

2  P. C. McIntyre and A. Fontcuberta i Morral, *Mater. Today Nano*, 2020, **9**, 100058.

3  K. A. Dick, K. Deppert, M. W. Larsson, T. Mårtensson, W. Seifert, L. R. Wallenberg and L. Samuelson, *Nat. Mater.*, 2004, **3**, 380–384.

4  L. Güniat, P. Caroff and A. Fontcuberta i Morral, *Chem. Rev.*, 2019, **119**, 8958–8971.

5  M. S. Gudiksen, L. J. Lauhon, J. Wang, D. C. Smith and C. M. Lieber, *Nature*, 2002, **415**, 617–620.

6  K. A. Dick, C. Thelander, L. Samuelson and P. Caroff, *Nano Lett.*, 2010, **10**, 3494–3499.

7   P. Caroff, K. A. Dick, J. Johansson, M. E. Messing, K. Deppert and L. Samuelson, *Nat. Nanotechnol.*, 2009, **4**, 50–55.

8   R. E. Algra, M. A. Verheijen, M. T. Borgstrom, L.-F. Feiner, G. Immink, W. J. P. van Enckevort, E. Vlieg and E. P. A. M. Bakkers, *Nature*, 2008, **456**, 369–372.

9   G. Shen, P.-C. Chen, Y. Bando, D. Golberg and C. Zhou, *J. Phys. Chem. C*, 2008, **112**, 16405–16410.

10  D. Spirkoska, J. Arbiol, A. Gustafsson, S. Conesa-Boj, F. Glas, I. Zardo, M. Heigoldt, M. H. Gass, A. L. Bleloch, S. Estrade, M. Kaniber, J. Rossler, F. Peiro, J. R. Morante, G. Abstreiter, L. Samuelson and A. Fontcuberta i Morral, *Phys. Rev. B*, 2009, **80**, 245325.

11  M. De Luca, C. Fasolato, M. A. Verheijen, Y. Ren, M. Y. Swinkels, S. Kölling, E. P. A. M. Bakkers, R. Rurali, X. Cartoixà and I. Zardo, *Nano Lett.*, 2019, **19**, 4702–4711.

12  Z. Ikonic, G. P. Srivastava and J. C. Inkson, *Phys. Rev. B*, 1995, **52**, 14078–14085.

13  Z. Ikonic, G. P. Srivastava and J. C. Inkson, *Phys. Rev. B*, 1993, **48**, 17181–17193.

14  M. Y. Swinkels and I. Zardo, *J. Phys. D*, 2018, **51**, 353001.

15  N. Guan, X. Dai, A. V. Babichev, F. H. Julien and M. Tchernycheva, *Chem. Sci.*, 2017, **8**, 7904–7911.

16  P. Krogstrup, H. I. Jørgensen, M. Heiss, O. Demichel, J. V. Holm, M. Aagesen, J. Nygard and A. Fontcuberta i Morral, *Nat. Photonics*, 2013, **7**, 306–310.

17  S. A. Mann, R. R. Grote, R. M. Osgood Jr., A. Alu and E. C. Garneet, *ACS Nano*, 2016, **10**, 8620–8631.

18  Z. Ikonic, G. P. Srivastava and J. C. Inkson, *Solid State Commun.*, 1993, **86**, 799–802.

19  X.-S. Fang, C.-H. Ye, L.-D. Zhang and T. Xie, *Adv. Mater.*, 2005, **17**, 1661–1665.

20  Q. Li, X. Gong, C. Wang, J. Wang, K. Ip and S. Hark, *Adv. Mater.*, 2004, **16**, 1436–1440.

21  G. Shen, Y. Bando, B. Liu, C. Tang and D. Golberg, *J. Phys. Chem. B*, 2006, **110**, 20129–20132.

22  R. Wagner and W. Ellis, *Applied Physics Letters*, 1964, **4**, 89–90.

23  R. Katsube and Y. Nose, *J. Solid State Chem.*, 2019, **280**, 120983.

24  S. Escobar Steinvall, N. Tappy, M. Ghasemi, R. R. Zamani, T. LaGrange, E. Z. Stutz, J.-B. Leran, M. Zamani, R. Paul and A. Fontcuberta i Morral, *Nanoscale Horiz.*, 2020, **5**, 274–282.

25  J. P. Bosco, G. M. Kimball, N. S. Lewis and H. A. Atwater, *J. Cryst. Growth*, 2013, **363**, 205–210.

26  R. Paul, N. Humblot, S. E. Steinvall, E. Z. Stutz, S. S. Joglekar, J.-B. Leran, M. Zamani, C. Cayron, R. Logé, A. G. del Aguila, Q. Xiong and A. F. i Morral, *Cryst. Growth Des.*, 2020, 3816–3825.

27  N. Y. Dzade, *Phys. Chem. Chem. Phys.*, 2020, **22**, 1444–1456.

28  M. Y. Swinkels, A. Campo, D. Vakulov, W. Kim, L. Gagliano, S. Escobar Steinvall, H. Detz, M. De Luca, A. Lugstein, E. P. A. M. Bakkers, A. Fontcuberta i Morral and I. Zardo, *Phys. Rev. Appl.*

29  M. V. Stackelberg and R. Paulus, *Z. phys. Chem*, 1935, **28**, 427–460.

30  M. Bhushan and A. Catalano, *Appl. Phys. Lett.*, 1981, **38**, 39–41.

31  T. Suda, K. Kakishita, H. Sato and K. Sasaki, *Appl. Phys. Lett.*, 1996, **69**, 2426–2428.

32  H. S. Im, K. Park, D. M. Jang, C. S. Jung, J. Park, S. J. Yoo and J.-G. Kim, *Nano Lett.*, 2015, **15**, 990–997.

33  H. S. Kim, Y. Myung, Y. J. Cho, D. M. Jang, C. S. Jung, J. Park and J.-P. Ahn, *Nano Lett.*, 2010, **10**, 1682–1691.

34  S. B. Choi, M. S. Song and Y. Kim, *J. Phys. Chem. C*, 2019, **123**, 4597–4604.

35  W. Shockley and H. J. Queisser, *J. Appl. Phys.*, 1961, **32**, 510–519.

36  S. Rühle, *Sol. Energy*, 2016, **130**, 139–147.

37  R. Yang, Y.-L. Chueh, J. R. Morber, R. Snyder, L.-J. Chou and Z. L. Wang, *Nano Lett.*, 2007, **7**, 269–275.

38  G. Lombardi, F. de Oliveira, M. Teodoro and A. Chiquito, *Appl. Phys. Lett.*, 2018, **112**, 193103.

39  K. A. Brakke, *Exp. Math.*, 1992, **1**, 141–165.

40  *MSA for Digital Micrograph*, HREM Research, https://www.hremresearch.com/Eng/plugin/MSAEng.html.

41  F. de la Pena, T. Ostasevicius, V. Tonaas Fauske, P. Burdet, P. Jokubauskas, M. Nord, M. Sarahan, E. Prestat, D. N. Johnstone, J. Taillon, undefined Jan Caron, T. Furnival, K. E. MacArthur, A. Eljarrat, S. Mazzucco, V. Migunov, T. Aarholt, M. Walls, F. Winkler, G. Donval, B. Martineau, A. Garmannslund, L.-F. Zagonel and I. Iyengar, *Microsc. Microanal.*, 2017, **23**, 214–215.

42  J. Mooney and P. Kambhampati, *J. Phys. Chem. Lett.*, 2014, **5**, 3497–3497.

43  C. Klein, B. Dutrow, J. D. Dana and C. Klein, *Manual of mineral science*, Wiley New York, 2002.

44  R. R. Zamani, F. S. Hage, S. Lehmann, Q. M. Ramasse and K. A. Dick, *Nano Lett.*, 2018, **18**, 1557–1563.

45  J. A. Mundy, Q. Mao, C. M. Brooks, D. G. Schlom and D. A. Muller, *Appl. Phys. Lett.*, 2012, **101**, 042907.

46  D. Bhattacharyya, X.-Y. Liu, A. Genc, H. L. Fraser, R. G. Hoagland and A. Misra, *Appl. Phys. Lett.*, 2010, **96**, 093113.

47  D. Bhattacharyya, N. A. Mara, P. Dickerson, R. G. Hoagland and A. Misra, *Acta Mater.*, 2011, **59**, 3804–3816.

48  R. R. Zamani and J. Arbiol, *Nanotechnology*, 2019, **30**, 262001.

49  A. P. Goldstein, S. C. Andrews, R. F. Berger, V. R. Radmilovic, J. B. Neaton and P. Yang, *ACS Nano*, 2013, **7**, 10747–10751.

50  J. Hoemke, E. Tochigi, T. Tohei, H. Yoshida, N. Shibata, Y. Ikuhara and Y. Sakka, *J. Am. Ceram. Soc*, 2018, **101**, 2616–2626.

51  J. Hoemke, E. Tochigi, T. Tohei, H. Yoshida, N. Shibata, Y. Ikuhara and Y. Sakka, *J. Am. Ceram. Soc*, 2017, **100**, 4252–4262.

52  H. Schmid, E. Okunishi, T. Oikawa and W. Mader, *Micron*, 2012, **43**, 49–56.

53  T. Burgess, S. Breuer, P. Caroff, J. Wong-Leung, Q. Gao, H. Hoe Tan and C. Jagadish, *ACS Nano*, 2013, **7**, 8105–8114.

54  N. Isik Goktas, A. Sokolovskii, V. G. Dubrovskii and R. R. LaPierre, *Nano Lett.*, 2020, **20**, 3344–3351.

55  M. de la Mata, R. R. Zamani, S. Martí-Sánchez, M. Eickhoff, Q. Xiong, A. Fontcuberta i Morral, P. Caroff and J. Arbiol, *Nano Lett.*, 2019, **19**, 3396–3408.

56  L. Ghisalberti, H. Potts, M. Friedl, M. Zamani, L. Güniat, G. Tütüncüoglu, W. C. Carter and A. F. i Morral, *Nanotechnology*, 2019, **30**, 285604.

57  F. M. Ross, J. Tersoff and M. C. Reuter, *Phys. Rev. Lett.*, 2005, **95**, 146104.

58  G. M. Kimball, A. M. Mueller, N. S. Lewis and H. A. Atwater, *Appl. Phys. Lett.*, 2009, **95**, 112103.

59  R. Katsube, H. Hayashi, A. Nagaoka, K. Yoshino, Y. Nose and Y. Shirai, *Jpn. J. Appl. Phys*, 2016, **55**, 041201.

60  J. Andrzejewski and J. Misiewicz, *Phys. Status Solidi B*, 2001, **227**, 515–540.

# Supplementary Information to "Heterotwin $Zn_3P_2$ superlattice nanowires: the role of indium insertion in the superlattice formation mechanism and their optical properties."

Simon Escobar Steinvall[1*], Lea Ghisalberti[1*], Reza R. Zamani[2*], Nicolas Tappy[1], Fredrik S. Hage[3], Elias Stutz[1], Mahdi Zamani[1], Rajrupa Paul[1], Jean-Baptiste Leran[1], Quentin M. Ramasse[3], W. Craig Carter[1,4], Anna Fontcuberta i Morral[1,5]

* Equal contribution

1 Laboratory of Semiconductor Materials, Institute of Materials, École Polytechnique Fédérale de Lausanne, 1015 Lausanne, Switzerland

2 Centre Inderdisciplinaire de Microscopie Électronique, École Polytechnique Fédérale de Lausanne, 1015 Lausanne, Switzerland

3 SuperSTEM Laboratory, SciTech Daresbury Campus, Keckwick Lane, Warrington WA4 4AD, United Kingdom

4 Department of Materials Science and Engineering, Massachusetts Institute of Technology, Cambridge, Massachusetts 02139, USA

5 Institute of Physics, École Polytechnique Fédérale de Lausanne, 1015 Lausanne, Switzerland

## Parameter definition

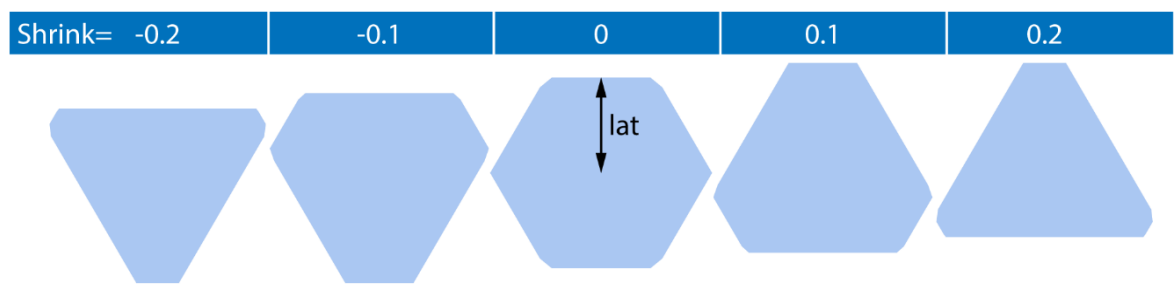

**Figure S1.** Illustration of the parameters used to model the droplet behaviour.

## Core-loss electron energy-loss spectra

Fitting of the EEL maps are done using the peak at 443 eV for In and 1020 eV for Zn. EEL spectra after denoising are shown in Figure S2, showing the In (a) and Zn signal (b). The fitting is done using the software Gatan Digital Micrograph v2.32.

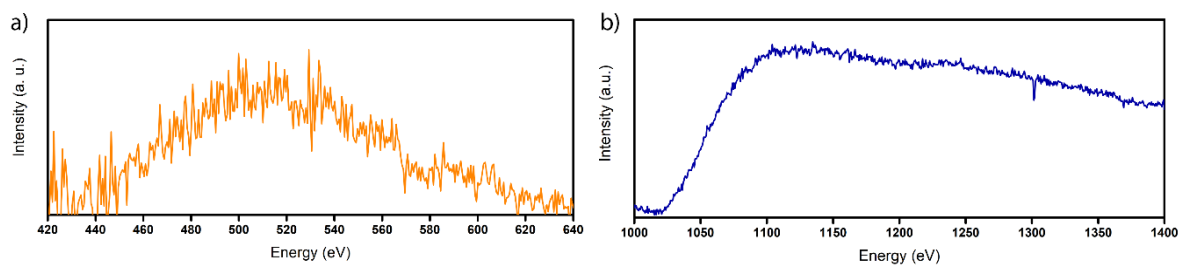

**Figure S2.** (a) EEL spectrum showing the In peak and (b) EEL spectrum showing the Zn peak used for mapping.

## Oscillation Modelling

In Figure S3a we show a HR-TEM image of a zigzag $Zn_3P_2$ nanowire. We can clearly observe the presence of a sharp interface between regions having different crystal orientation (having different contrast as well). This interface occurs where the heterotwin forms and produces the change the crystal orientation. To investigate the driving forces at the origin of this growth process, we looked for a periodicity rule in the insertion of these heterotwins as a function of the length of the nanowire. Due to the tapering effect influencing the width of the nanowire along its growth axis, we decided to measure the position and the nanowire's width at which the heterotwin is inserted through the software ImageJ. In Figure S3b we show the data points collected and the periodic function fitting the data.

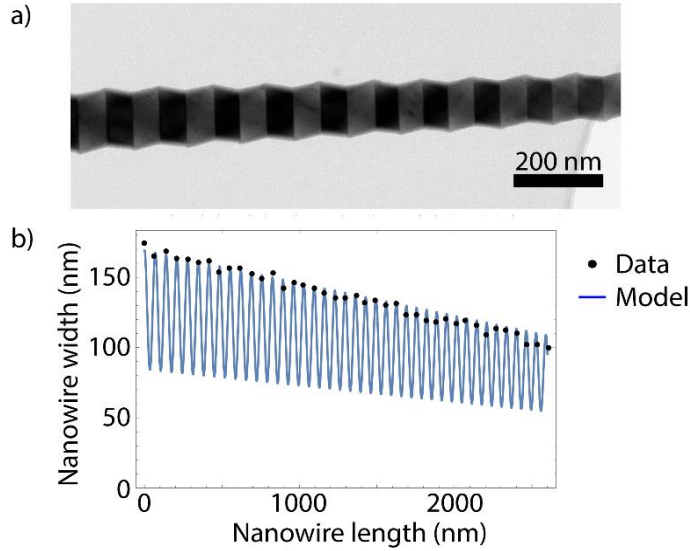

**Figure S3.** a) TEM image of zigzag $Zn_3P_2$ nanowire grown by MBE and b) plot of the development of an approximation of the nanowire's width as a function of the nanowire's growth axis: the dots represent the measurements taken on the sample shown in a) with fitting function reported in equation (1).

The selected periodic function (eq. 1) has a linear decay in the amplitude and an exponential decay in the frequency of insertion of the heterotwin. $W_o$ and $h_o$ are the initial width of the nanowire and the initial separation between the first two consecutive heterotwins observed in the nanowire. The fitting to the data produces $\pi/2.02$ (89°) for the tapering factor, which corresponds to the tapering angle measured in zigzag nanowires through SEM analysis (i.e., the amplitude linear decay); and $2 \times 10^{-5}$ for the continuous decay rate of the separation, characterizing the exponential decay in the heterotwin insertion periodicity. We believe that the tapering factor and the continuous decay rate coefficient depends on the MBE growth conditions, i.e. temperature and II-V ratio.

$$W(x) = \left(W_0 - \frac{W_0 - \frac{W_0}{2}}{2} - \frac{x}{\tan\frac{\pi}{2.02}}\right)\left(1 + \frac{1}{3}\cos\frac{2\pi x}{h_0 e^{-2\times10^{-5}x}}\right) \quad (1)$$

**Casino simulations of energy deposition in a superlattice nanowire**

The simulations performed using CASINO software (V3.3) are presented in Figure S4. It can be observed that the extent of the beam interaction volume in $Zn_3P_2$ does not exceed 50 nm in depth and laterally at an acceleration voltage of 3kV. The assumption that no energy reaches through the sample at this acceleration voltage is thus well verified for most probed points on the CL map. Additionally, the total amount of energy deposited in the sample varies by 17% between outwards and inwards facing apices. This is explained by the variation in backscattering coefficient ($\eta$), which is largely influenced by the local geometry. We make the argument that the CL emission should follow the energy deposited by the beam. Accordingly, the superlattice nanowire local geometry is sufficient to explain the dark edge-contrast observed in the panchromatic CL map (Figure 3).

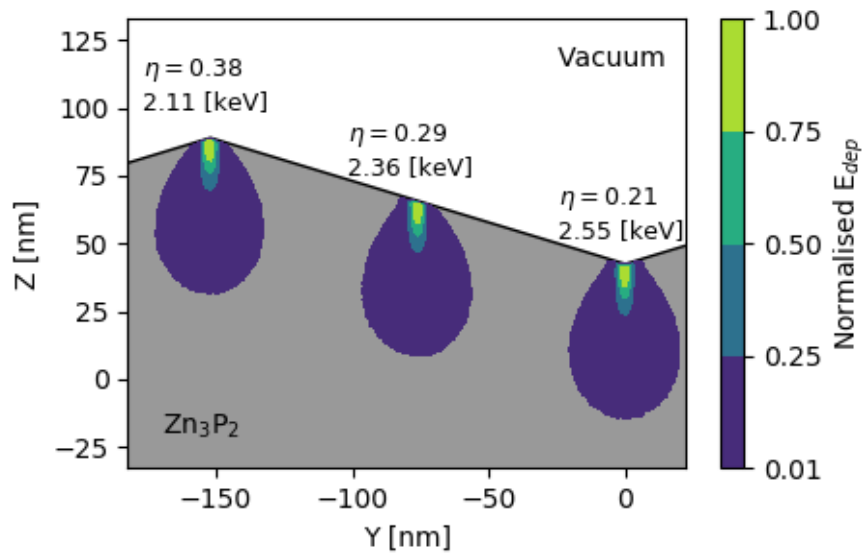

**Figure S4** CASINO Simulation of energy deposited in the sample for different edge configuration. Coloured surfaces show the decrease in energy density deposited in the sample, normalised to the maximum. Annotations show the fraction of backscattered electrons (η) and the total amount of energy (in keV) deposited in the sample per electron, for each edge configuration. It is observed that outwards facing apices (left) show enhanced backscattering compared to inwards facing apices (right) or facets (middle). Simulations where performed using a collimated electron beam of 10 nm diameter at 3 kV and a density of 4.55 g cm$^{-3}$ for $Zn_3P_2$. Due to the limited possibilities of simulating complex geometries in CASINO 3, we model the sample with truncated pyramids. This approximation reproduces well local edge configuration, but would not be valid in experiment conditions where a significant part of the beam energy is transmitted through the sample, e.g. at high beam energies.